\documentclass[sigplan]{acmart}

\setcopyright{none}

\usepackage{booktabs}   
\usepackage{subcaption} 
\usepackage[linesnumbered,ruled,lined]{algorithm2e}               
\usepackage{epstopdf}
\usepackage{multirow}
\usepackage{listings}
\newcommand{\ie}{{\it i.e.,}}
\newcommand{\eg}{{\it e.g.,}}

\begin{document}

\title[Heartbeat Diagnosis in OpenMP Multi-Threaded Systems]{Heartbeat Diagnosis of Performance Anomaly in OpenMP Multi-Threaded Systems}         


\author{Weidong Wang}

\orcid{0000-0002-7378-2766}             
\affiliation{
  \department{Faculty of Information Technology}              
  \institution{Beijing University of Technology}            
  \streetaddress{Chaoyang District Pingleyuan No.100}
  \city{Beijing}
  \state{}
  \postcode{100124}
  \country{China}                    
}
\email{wangweidong@bjut.edu.cn}          

\author{Wangda Luo}
     
\orcid{nnnn-nnnn-nnnn-nnnn}             
\affiliation{
  \department{Faculty of Information Technology}             
  \institution{Beijing University of Technology}           
  \streetaddress{}
  \city{Beijing}
  \state{}
  \postcode{}
  \country{China}                   
}
\email{luowangda_bjut@163.com}         

\begin{abstract}
This paper presents a novel heartbeat diagnosis regarding performance anomaly for OpenMP multi-threaded applications. First, we design injected heartbeat APIs for OpenMP multi-threaded applications. Then, we leverage the heartbeat sequences to extract features of previously-observed anomalies. Finally, we adopt a tree- based algorithm, namely HSA, to identify the features that are required to diagnose anomalies. To evaluate our framework, the NAS Parallel NPB benchmark, EPCC OpenMP micro-benchmark suite, and Jacobi benchmark are used to test the performance of our approach proposed. 

\end{abstract}



\maketitle

\section{Introduction}

In the traditional field of computer research, such as underlying architecture \cite{Marongiu2012,Yamazaki2018} and multiple threads \cite{Aldea2016} based software design, the study heartbeats in a multi-threaded program has gradually become a hot spot in both industry and academia. According to the analysis of heart rate, we can better understand the information of execution state and exception, and provide a guarantee for program running reliability. While many techniques have been proposed for detecting the root cause of anomalies \cite{Hiranya2017} in OpenMP applications, these techniques still rely on human directors to identify the root causes of the anomalies, leading to wasted computing resources. An effective way of decreasing the impact of anomalies is to automate the diagnosis of anomalies \cite{Yu2016}, which lays the foundation for automated diagnosis. Our contribution is as follows.

\begin{itemize}
	\item First, we propose a low-overhead heartbeat based anomaly detection framework of multi-threaded OpenMP applications that enables detection and diagnosis of previously observed anomalies. To simplify framework use, we design heartbeat API functions for portable use.
	\item Second, we propose a hybrid heartbeat sequence analysis algorithm for anomaly diagnosis, namely HSA, in which we design an easy-to-compute heartbeat features (\eg\ completion time ratio and heart rate ratio) and employ the DTW distance and LB\_Keogh to detect target anomalies.
\end{itemize}

\section{Heartbeat APIs for OpenMP Applications}
\label{sec:Diagnosis}

\begin{figure}[h]
	\centering
	\includegraphics[width=9cm]{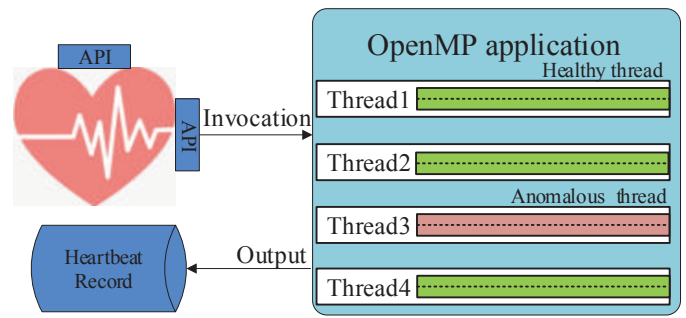}
	\caption{Heartbeat API for an OpenMP application.}\label{fig:EMF}
\end{figure}

To make any OpenMP application detected, we design a heartbeat generator that can generate a heartbeat time series. Since the design of the heartbeat generator must be efficient and simple as it is required, easy to use for most programmers. According to the characters of OpenMP applications, we design a group of heartbeat APIs, which can be invoked by an OpenMP application as shown in Figure \ref{fig:EMF}. 

The OpenMP application embedded heartbeat APIs can periodically output heartbeat timestamps to the specified document. In Table \ref{FUNC}, we list three heartbeat functions. \emph{Heartbeat\_OpenMP\_Init} denotes the initialization function, which takes charge of initializing variables and allocating memory related with heartbeat for all OpenMP threads. \emph{Heartbeat\_OpenMP\_Generate} represents the generation function that can periodically output heartbeats. \emph{Heartbeat\_OpenMP}\\ \emph{\_Finished} is the finish function and it is responsible for heartbeat end and flush memory.

Here, we illustrate an example to explain how to generate heartbeats in an OpenMP application as in List 1 as follows.

\lstset{language=c, caption={An example of heartbeat API invoked by an OpenMP application.}, label=inscon,
	basicstyle=\footnotesize\ttfamily,captionpos=b,escapeinside=``}


\begin{lstlisting}{t}
#`\textbf{include}` "Heartbeat_Support_OpenMP.h"
#`\textbf{define}` NUM_THREADS 4
int `\textbf{main}`(int argc, char **argv){
 L_Heartbeat_OpenMP_Init();  
 #`\textbf{pragma omp parallel for}`
 `\textbf{for}` (iter = 1; iter < n; j++){
   threadnum = omp_get_thread_num();
   L_Heartbeat_OpenMP_Generate(threadnum,1,iter); 
   }
 Heartbeat_OpenMP_Finished();
 return(0);
}
\end{lstlisting}

For the OpenMP application specified, the initialization function, \emph{Heartbeat\_OpenMP\_Init}, should firstly be injected by the OpenMP main process or thread. Then, the generation function \emph{Heartbeat\_OpenMP\_Generate} is invoked when executing the for loop by each OpenMP thread. Therefore, a heartbeat automatically generated with sequence number, current time, and thread ID at each time the loop is taken. In this way, we can determine the latency and further judge the performance anomalies of these working threads by comparing them with the anticipated heartbeat sequence. Finally, \emph{Heartbeat\_OpenMP\_Finished} is executed to stop the heartbeat and flush memory.

\section{Heartbeat Diagnosis}
\label{subsec:MT}

Our heartbeat diagnosis is based on a subset of heartbeat sequence and we calculate the statistical features of the heartbeat sequence. For each collected heartbeat sequence, we keep track of the recent window size observed in a sliding window, and calculate the following statistical features such as \textit{GlobalTimeRatio} and \textit{GlobalHeartbeatRatio}. To enable easy scaling, we extract these statistical features from individual working threads and do not account for the interaction and correlation among multiple threads. With a constant and small window size, this enables us to generate features at runtime with negligible overhead. These features are derived from the heartbeat sequence that is determined offline based on the target anomalies and the target OpenMP applications. 

\subsection{Heartbeat Sequence Pre-processing}
\label{subsec:NDA}

\begin{figure}[bhtp]
	\centering
	\subcaptionbox{\label{1}}{\includegraphics[width=4cm]{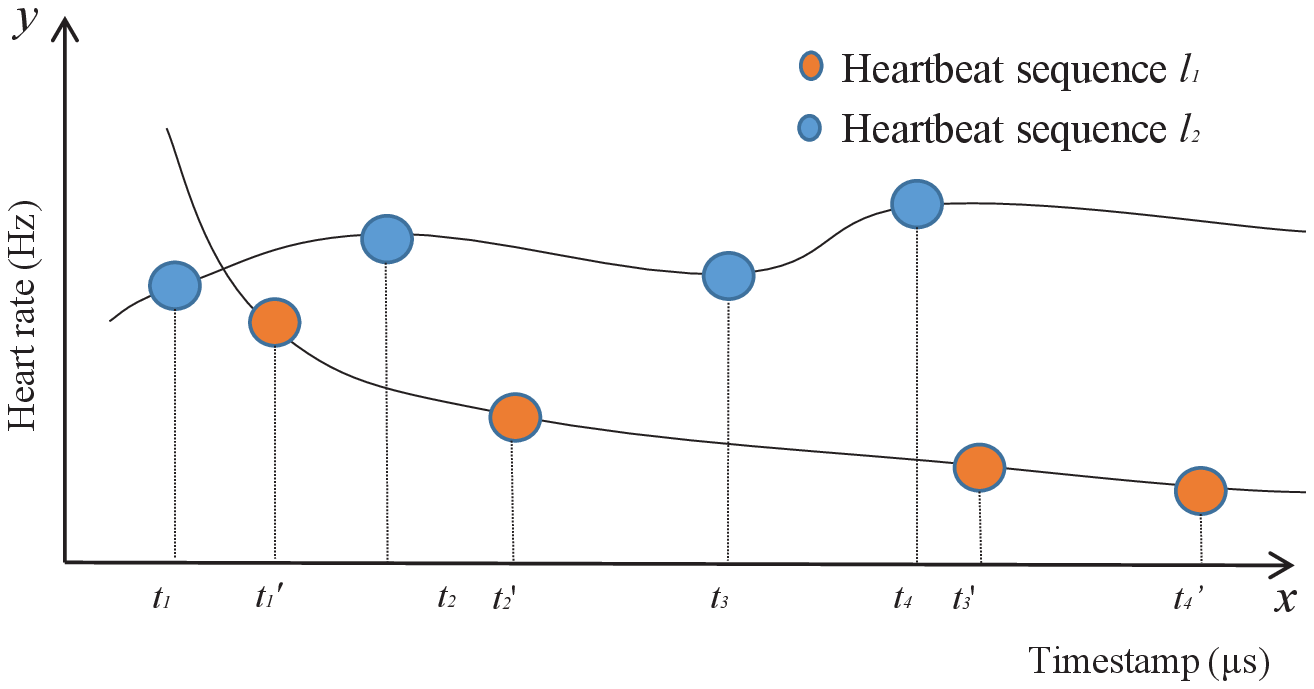}}\hfill
	\subcaptionbox{\label{2}}{\includegraphics[width=4cm]{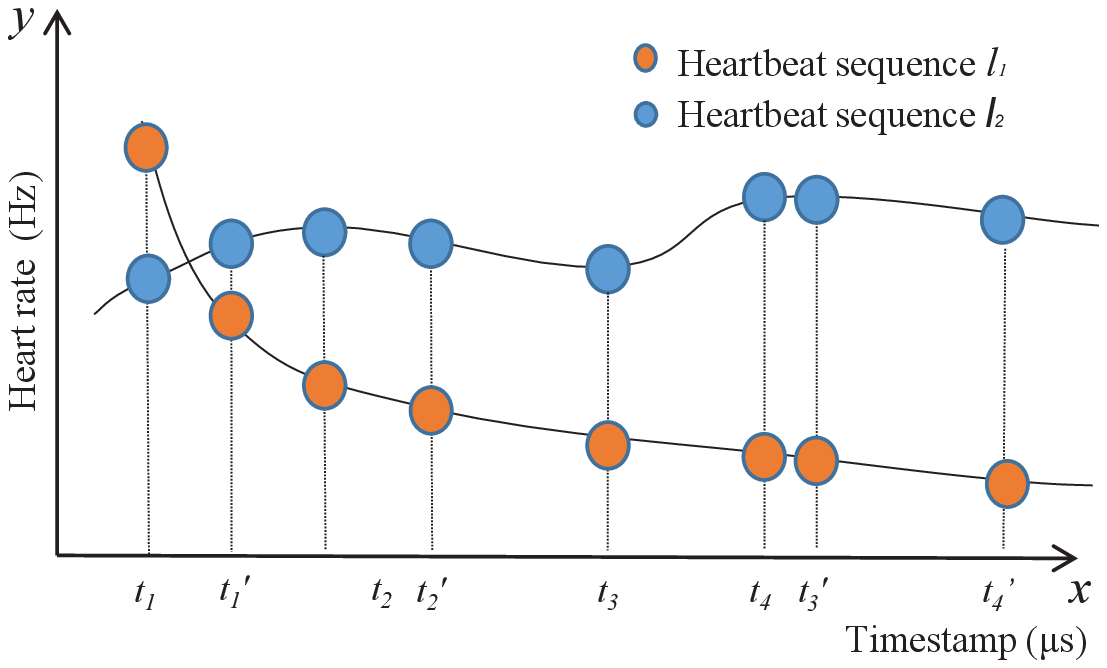}}
	\caption{Heartbeat sequence alignment by polynomial regression.}\label{fig-5}
\end{figure}

To spot the dubious data, we first observe the timestamp at the specified window size $w'$. So we scale timestamp at the window size shown in the $x$-axis and observe their heart rate shown in the y-axis. Figure \ref{fig-5} shows both dubious heartbeat sequence fragment $l_1$ and heartbeat sequence fragment $l_2$, respectively. In a general sense, it can be observed that the fragment $l_1$ always fluctuates around the convergence value of heart rate, while the fragment $l_2$ gradually decreases.

For better studying the relationship of heartbeat sequence fragment $l_1$ and $l_2$, we commonly compare the two sequences that have different timestamps. First, we align two arbitrary timestamps of the sequence fragment $l_1$ and $l_2$. To obtain the horizontal alignment, we fix the one of $l_1$ and $l_2$, then use polynomial regression to fit the scatter points of the fixed one. In this way, we can obtain a continuous curve related to the fixed sequence in Equation \ref{formu1} as follows.

\begin{equation}
	\footnotesize
	y = b+w_1x+x_2x^2+w_3x^3+\cdots+w_nx^n,
	\label{formu1}
\end{equation}

\noindent where $\omega$ = [$\omega_1,\omega_2,\cdots,\omega_n$] denotes the target weight; and $b$ is the target $y$-intercept. To obtain the values of two parameters $\omega$ and $b$, we use the given heartbeat sequences $(X,Y)$ = [($x_1,y_1$),($x_2,y_2$),$\cdots$,($x_n,y_n$)] as the training samples. The optimized objective function is as follows.

\begin{equation}
	\footnotesize
	\mathop{\arg\min}_{\omega,b} {L(w,b)} = \sum_{i=1}^n(y_i-w_0-w_1x_i-w_2x_i^2-\cdots-w_nx_i^n)^2,
	\label{formu2}
\end{equation}

\noindent where $L(w,b)$ also call the loss function. To evaluate the polynomial regression, we employ R-squared which is a statistical measure of how close the actual value is to the fitted regression results. The R-squared is as follows.

\begin{equation}
	\footnotesize
	R-squared = 1-\frac{\sum\limits_{i=1}^n (\hat{y_i}-y_i)^2}{\sum\limits_{i=1}^n (\bar{y}-y_i)^2},
	\label{formu3}
\end{equation}

\noindent where $\hat{y_i}$ denotes the prediction value and $\bar{y}$ is the average value. R-squared value is between 0 and 1. The closer it is to 1, the better the regression fitting effect will be. To obtain a better fitting result, we try to increase the degree of polynomial. When the R-squared value obtained is larger than 0.9, such a polynomial fitting result can be accepted. For instance, as shown in Figure \ref{fig-5}(a), we obtain two continuous heart rate through polynomial regression. Then, we calculate the heart rate on different curves at the same timestamp. Figure \ref{fig-5}(b) is the result of alignment.  

\subsection{Feature Extraction}
\label{subsec:FE}

To judge whether a heartbeat sequence differs from a normal sequence, we define the basic features of a heartbeat sequence based on heart rate, completion time, lower bound, and similarity.

\textbf{Step 1}: Suppose there are two sequences $Q$ and $C$ as shown in Equation \ref{formu11}. And the length is $m$ and $n$, respectively.

\begin{equation}
	\footnotesize
	\begin{aligned}
		& Q={(t_1,q_1),(t_2,q_2),(t_3,q_3),\cdots,(t_n,q_n)} \\
		& C={(t'_1,c_1),(t'_2,c_2),(t'_3,c_3),\cdots,(t'_m,c_m)}
	\end{aligned}
	\label{formu11}
\end{equation}

To measure the ratio of completion time, the global time ratio of two heartbeat sequences is defined in Equation \ref{formu132}.

\begin{equation}
	\footnotesize
	GlobalTimeRatio(C,Q)=\frac{t'_m}{t_n},
	\label{formu132}
\end{equation}

\noindent where $t'_m$ is the completion time of heartbeat sequence $C$, and $t_n$ is the completion time of heartbeat sequence $Q$. To measure the local differences of completion time, we employ a local fine-grained time ratio based on the sliding window as follows.

\begin{equation}
	\footnotesize
	LocalTimeRatio(C,Q,w,k)=\frac{\sum\limits_{i=1}^n \frac{t'_{i+w}-t'_{i}}{t_{i+w}-t_{i}}}{k},
\end{equation}

\noindent where $t'_{i+w}$ denotes the completion time of sequence $C$ at the ($i+w$)-th timestamp. And $t_{i+w}$ denotes the completion time of sequence $Q$ at the ($i+w$)-th timestamp. $w$ is the size of sliding window. $k$ represents the number of sliding windows.

To measure the ratio of heart rate between two heartbeat sequences, in this paper the heartbeat ratio is defined in Equation (\ref{formu133}).

\begin{equation}
	\footnotesize
	GlobalHeartbeatRatio(C,Q)=\frac{\frac{\sum\limits_{i=1}^mc_i}{m}}{\frac{\sum\limits_{i=1}^nq_i}{n}},
	\label{formu133}
\end{equation}

\noindent where $m$ and $n$ represent the length of heartbeat sequence $C$ and $Q$. $c_i$ and $q_i$ are the heart rate of $i$-th heartbeat belong to sequence $C$ and $Q$. To measure the local differences of heartbeat ratio, we employ a local fine-grained heartbeat ratio based on the sliding window as follows.

\begin{equation}
	\footnotesize
	LocalHeatbeatRatio(C,Q,w,k)=\frac{\sum\limits_{i=1}^n \frac{c_{i+w}-c_{i}}{q_{i+w}-q_{i}}}{k},
	\label{formu1355}
\end{equation}

\noindent where $c_{i+r}$ denotes the heart rate of sequence $C$ at the ($i+w$)-th timestamp. And $t_{i+w}$ denotes the heart rate of sequence $Q$ at the ($i+w$)-th timestamp. $w$ represents the number of sliding windows. $k$ is the size of sliding window.

To measure a similarity between two heartbeat sequences, DTW, dynamic time wrapping \cite{ding2013developments} is employed for sequence analysis. The DTW distance can be found by an optimal bending path to minimize the cumulative distance of two heartbeat sequences. 

\textbf{Step 2}: The DTW distance is calculated between two heartbeat sequences as follows.

\begin{equation}
	\footnotesize
	DTW(q_i,c_j)=\sum\limits_{i=0}^m \sum\limits_{j=0}^n D(q_i,c_j),
	\label{formu6}
\end{equation}

\noindent where $D(q_i,c_j)$ is the dynamic time warping function, which can obtain an optimal match between two given heartbeat sequences with certain restrictions and rules as follows.

\begin{equation}
	\footnotesize
	\left\{
	\begin{aligned}
		&0, (i=0,j=0) \\
		&\left|q_i-c_j\right|+D(q_{i-1},c_j),(i\geq 1,j=0) \\
		&\left|q_i-c_j\right|+D(q_{i},c_{j-1}), (i=0,j\geq 1)\\
		&\left|q_i-c_j\right|+\min[D(q_{i-1},c_{j-1}),D(q_{i-1},c_j),D(q_i,c_{j-1})], i,j \geq 1\\
	\end{aligned}
	\right.
	\label{formu6-7}
\end{equation}

\noindent where $|q_i-c_j|$ represents Euclidean distance between the heart rate of $i$-th heartbeat in the sequence $C$ and the heart rate of $j$-th heartbeat in the sequence $Q$. The optimal match is denoted by the match that satisfies all the restrictions above and that has the minimal cost, which is calculated as the sum of absolute differences for each matched pair of indices between their values. Hence, the minimal cost distance can be easily calculated by Dynamic Programming \cite{bertsekas1995dynamic}. For the sake of processing speed, as the complement of DTW distance, we also introduce the LB\_Keogh lower bound \cite{keogh2006lb_keogh} since it can filter most of the sequences that cannot be the optimal matching heartbeat sequences as follows.

\textbf{Step 3}: LB\_Keogh lower bound is introduced between two heartbeat sequences.

First, we define upper bound sequence $u=u_1,u_2,\cdots,u_n$ for the heartbeat sequence $Q=q_1,q_2,\cdots\,q_n$ as follows.

\begin{equation}
	\footnotesize
	\left\{
	\begin{aligned}
		&u_i = \max(q_0,q_1,\cdots,q_{i+w}), i<w \\
		&u_i = \max(q_{i-w},q_{i-w+1},\cdots,q_{i+w}), i\geq w \\
	\end{aligned}
	\right.
	\label{formu10}
\end{equation}

Similarly, we also define lower bound sequence $l=l_1,l_2,\cdots,l_n$ for the heartbeat sequence $Q=q_1,q_2,\cdots\,q_n$ as follows.

\begin{equation}
	\footnotesize
	\left\{
	\begin{aligned}
		&l_i = \min(q_0,q_1,\cdots,q_{i+w}), i<w \\
		&l_i = \min(q_{i-w},q_{i-w+1},\cdots,q_{i+w}), i\geq w \\
	\end{aligned}
	\right.
	\label{formu101}
\end{equation}

\noindent where $w$ denotes the size of the sliding window. Hence, for the given heartbeat sequence $Q$, we can obtain two sequences, \ie\ upper bound heartbeat sequence $u$ and lower bound sequence $l$.

Then, we calculate the LB\_Keogh value between the given heartbeat sequence $Q$ and the target heartbeat sequence $C$ in Equation \ref{formu10}.

\begin{equation}
	\footnotesize
	LB\_Keogh(Q,C)=\sum\limits_{i=1}^n \left\{
	\begin{aligned}
		&(c_i-u_i)^2,c_i>u_i \\
		&(c_i-l_i)^2,c_i<u_i\\
		&0, otherwise
	\end{aligned}
	\right.
	\label{formu10}
\end{equation}

\noindent where $c_i$ denotes $i$-th heartbeat data in the target heartbeat sequence $C=c_1,c_2,\cdots,c_n$. The accumulative differences are calculated by Euler distance among the upper bound $u$ and lower bound sequence $l$ of the given heartbeat sequence $Q$.

\subsection{Heartbeat Diagnosis Algorithm}
\label{subsec:FE}

The goal of the algorithm is to discover anomalous threads of OpenMP applications by heartbeat diagnosis. First, we use the normal range of those features defined in the previous section. Inspired by the decision tree algorithm that introduces the tree structure logic judgment method to diagnose the current state using the heartbeat data collected, we propose a HAS algorithm as our heartbeat diagnosis of performance anomaly. The HAS algorithm adopts a white-box parameter training method for the sake of good interpretability. Also different from the traditional greedy based decision tree algorithm, which only illustrates that the local optimal solution is sought at each tree node yet the global optimal solution may be ignored, we employ the confidence interval of each heartbeat feature as the tree node to guarantee the global or near-global optimal solution. The HAS algorithm is as follows. 

\begin{algorithm}
	\caption{Heartbeat diagnosis of performance anomaly}
	\LinesNumbered
	\KwIn{The features of a group of heartbeat sequences: \newline
		\{$GTR,LTR,GHR,LHR,DTW,LB$\}; \newline
		Normal range of heartbeat sequence: \{$DistanceRange,TimeRange,HeartrateRange$\};}
	\KwOut{Each heartbeat sequence's status in the group:\newline
		$status[i]$;}
	\textbf{Initialize}: \newline
	$mintime$=min($TimeRange$); \newline
	$maxtime$=max($TimeRange$); \newline
	Enum $ThrStatus$=\{normal,memoryleak,shutdown\}; \newline
	\ForEach {$i \in$ heartbeat sequences}{
		$status[i]$= normal; \newline
		\eIf{$\{DTW_i,LB_i\} \subseteq DistanceRange$ == True}{
			\If{$\{GTR_i,LTR_i\} \subseteq TimeRange$ == False}{
				\If{$GTR_i<mintime$}{
					$status[i]$=shutdown;
				}
				\If{$GTR_i>maxtime$ \textbf{and} $\{GHR_i,LHR_i\} \subseteq HeartrateRange$ == False}{
					$status[i]$=memoryleak;
				}
			}
		}
		{
			$status[i]$=memoryleak; \newline
			\If{$GTR_i<mintime$}{
				$status[i]$=shutdown;
			}
		}
	}
	Return(status);
\end{algorithm}

The basic principle of the algorithm is the feature comparison between a target heartbeat sequence and a normal heartbeat sequence. The normal range of the features defined in the previous section can be obtained by training a group of heartbeat sequences. In Algorithm 1, $GTH$ and $LTR$ denote the GlobalTimeRatio and LocalTimeRatio, respectively. $GHR$ and $LHR$ are the GlobalHeartbeatRatio and LocalHeartbeatRatio. $DTW$ is the dynamic time warping distance. And LB represents the LB\_Keogh distance. The main idea of the algorithm is the judgment of \emph{normal, memory leak, shutdown} status of OpenMP applications. For example, if the $DTW$ and $LB\_Keogh$ distances are within the normal range, the result guarantees that the two heartbeat sequences are similar. Then, we compare such similar heartbeat sequences by analyzing $GTR$ and $LTR$, which describe the differences in completion time. Meanwhile, we also consider the change of $GHR$ and $LHR$, which represents the global and local heart rate. To some extent, the OpenMP programs with performance anomaly such as memory leak anomaly has a longer completion time and lower heart rate than those without performance anomaly, which is accorded by the following experiments. 

\section{Experimental Methodology}
\label{sec:experiments}

\subsection{OpenMP Benchmarks}
\label{subsec:benchmarks}

To evaluate our framework, we use a heartbeat dataset collected from representative \\OpenMP benchmark applications.  In our evaluation, we use three different types of OpenMP benchmarks \textit{i.e.} NPB, EPCC, and Jacobi. (1) The NAS Parallel Benchmarks (NPB) are widely utilized by the parallel computing community as a representative set of OpenMP applications. (2) The EPCC OpenMP micro-benchmark suite is developed by Edinburgh Parallel Computing Centre for measuring the overheads of synchronization, loop scheduling, and array operations in the OpenMP runtime library. These benchmarks run in multiple cores of various scientific workloads. And (3) The Jacobi is another scientific computing application for multiple system performance analysis.

We run parallel applications on four OpenMP threads, where the threads are utilized as much as possible using balancing workload per core. For each benchmark application, we synthesize heartbeats in the following section.

\subsubsection{Heartbeats Embedded}
\label{subsec:mc}

For each benchmark application, we analyze the source code and its description, and then insert heartbeat APIs into the source code. According to the requirement of heart rate, we consider the number of iterations and control the number of heartbeat generation. As shown in Table \ref{tab4-2}, we adjust the average heart rate by altering the number of iterations, where specified iterations can produce one heartbeat for reducing or increasing heart rate for the benchmark application specified.

\begin{table}[bhtp]
	\centering
	\caption{Heartbeats in different Benchmarks.}
	\begin{tabular}[width=7cm]{p{1.6cm}p{3.2cm}c}
		\toprule
		Benchmark                      & Heartbeat cycle                                      &	Heart rate (beats/s)                                   \\
		\midrule
		NPB-bt                      & Every 1 iteration                               &    97519.4                                              \\
		NPB-lu                      & Every 1 iteration                               &    346877.5                                             \\
		NPB-cg                      & Every 10 iterations                              &    176311.3                                             \\
		NPB-sp                      & Every 1000 iterations                            &    528756.4                                             \\
		Jacobi                      & Every 10000 iterations                           &    5488.1                                             \\
	    Arraybench                   & Every 1 iteration                               &    31695.64                                     \\
		\bottomrule
	\end{tabular}
	\label{tab4-2}
\end{table}

To make the heart rate of each benchmark application in the same order of magnitude as well, we commonly control iterations as much as possible by adjusting quantity iterations for one heartbeat generation. For example, we adjust the iterations of the benchmark application NPB-sp, \ie\ every 1000 iterations for one heartbeat.

\subsubsection{Heartbeat Dataset}
\label{subsec:dataset}

To evaluate our approach with the experiments controlled, we design synthetic anomalies that mimic commonly observed anomalies caused by application- or system-level issues. Furthermore, we use two anomaly scenarios: (1) memory leak anomalies, where the program with these anomalies executes during the entire application run, and (2) random-offset shutdown anomalies, where such anomalies starts at a randomly selected time while the application is running. 

To build a heartbeat dataset, we had established a synthetic heartbeat application for each famous OpenMP benchmark, including (1) the OpenMP applications with synthetic heartbeats of the NAS parallel benchmarks, \ie\ \emph{NPB sp, lu, bt, and cg}; (2) the synthetic heartbeat application of the micro EPCC benchmark, \ie\ \emph{EPCCArray}; and (3) the synthetic heartbeat application of OpenMP Jacobi used in scientific computing. In the event, a total of 540 thousand heartbeats were collected from different OpenMP benchmarks.

The heartbeat dataset provides a group of .xlsx files that record various heartbeats with the benchmark applications above. 

\subsection{Results}
\label{sec:Results}

In our experiments, we randomly select 30\%  the same size of training samples in the dataset while the remaining heartbeat data is used as testing samples. To avoid the experimental deviation by randomly selecting samples, we repeat the process above three times for each experiment, and take the average of experimental results. 

\subsection{Performance}
\label{sec:ADC}

To make sense of the process of anomaly diagnosis, we divided heartbeat samples into three groups, \ie\ normal group, memory leak group, and shutdown group. For each group, we test macro F-score. In Benchmark NPB-cg, the macro average F-score is 0.95.

\subsection{Overhead}
\label{sec:US}

The following experiment tests the overhead of the heartbeat method by using different benchmark applications including \emph{NPB seriers}, \emph{EPCC}, and \emph{Jacobi}. In the experiment, we insert heartbeat API into the above benchmark applications, and then record the change of CPU usage. The overhead in our paper is defined as shown in Equation \ref{formu:3}.

\begin{equation}
\footnotesize
Overhead=\frac{E_\alpha-E_\beta}{E_\beta} \\
\label{formu:3}
\end{equation}

Where $E_\alpha$ denotes CPU usage of a heartbeat-enabled benchmark application while $E_\beta$ denotes the CPU usage of benchmark application without heartbeats at runtime. The overall overhead is under 2.5\%. Note that, the x-axis denotes the magnitude of heartbeat frequency. Also, we study that the overheads of these benchmark applications have slightly different due to their basic CPU usage without heartbeats. 

\section{Conclusion}
\label{sec:con}

This paper solved the problem of how to diagnose performance anomaly for OpenMP multi-threaded applications

First, we proposed the low-overhead heartbeat-based anomaly diagnosis approach of multi-threaded OpenMP applications that enables automatic detection and diagnosis of previously observed anomalies. To simplify framework use, we designed the heartbeat APIs for portable use.

Second, we proposed the hybrid heartbeat-based analysis algorithm for anomaly diagnosis, namely HSA, in which we designed an easy-to-compute heartbeat features and employed the DTW distance and LB\_Keogh to diagnose the target anomalies. 

For the future work, we may consider several aspects to further improve the performance of the approach proposed.

\bibliography{Hierarchical_new}

\appendix

\end{document}